\documentclass[a4paper,11pt]{article}

\usepackage{aaskaiid}
\usepackage{hyperref}
\usepackage{xcolor}
\usepackage{enumitem}
\usepackage{orcidlink}

\title{Discovering and Characterising Exoplanets and Ultracool Dwarfs with the Square Kilometre Array}
\ShortTitle{Exoplanets and UCDs with the SKA}

\author[1,2]{Robert~D.~Kavanagh\orcidlink{0000-0002-1486-7188}}
\ShortName{Kavanagh et al.}
\author[3]{Juan~B.~Climent\orcidlink{0000-0002-5093-6208}}
\author[4,5]{Philippe~Zarka\orcidlink{0000-0003-1672-9878}}
\author[2,1]{Joseph~R.~Callingham\orcidlink{0000-0002-7167-1819}}
\author[6,7]{Yuka~Fujii\orcidlink{0000-0002-2786-0786}}
\author[4]{Corentin~K.~Louis\orcidlink{0000-0002-9552-8822}}
\author[8,4,5]{Laurent~Lamy\orcidlink{0000-0002-8428-1369}}
\author[9]{Mayank~Narang\orcidlink{0000-0002-0554-1151}}
\author[10]{J.~Sebastian~Pineda\orcidlink{0000-0002-4489-0135}}
\author[2,11]{Harish~K.~Vedantham\orcidlink{0000-0002-0872-181X}}
\author[2,11]{Sanne~Bloot\orcidlink{0000-0002-3601-6165}}
\author[12,5]{Jean-Mathias~Grie{\ss}meier\orcidlink{0000-0003-3362-7996}}
\author[3,13]{Jose~Carlos~Guirado\orcidlink{0000-0003-2722-1615}}
\author[14,15]{Simranpreet~Kaur\orcidlink{0000-0002-5820-2532}}
\author[9]{T.~Joseph~W.~Lazio}
\author[16]{Miguel~P\'erez-Torres\orcidlink{0000-0001-5654-0266}}
\author[17,18]{Alice~Zurlo\orcidlink{0000-0002-5903-8316}}

\affiliation[1]{Anton Pannekoek Institute for Astronomy, University of Amsterdam, 1098 XH Amsterdam, The Netherlands}
\emailAdd{r.d.kavanagh@uva.nl}
\affiliation[2]{ASTRON, The Netherlands Institute for Radio Astronomy, Oude Hoogeveensedijk 4, 7991 PD Dwingeloo, The Netherlands}
\affiliation[3]{Departament d’Astronomia i Astrof\'isica, Universitat de Val\`encia, C. Dr. Moliner 50, E-46100 Burjassot, Val\`encia, Spain}
\affiliation[4]{LIRA, Observatoire de Paris, Universit\'e PSL, Sorbonne Universit\'e, Universit\'e Paris Cit\'e, CY Cergy Paris Universit\'e, CNRS, 92190 Meudon, France}
\affiliation[5]{Observatoire Radioastronomique de Nançay, Observatoire de Paris, CNRS, PSL, Univ. Orl\'eans, Nançay, France}
\affiliation[6]{National Astronomical Observatory of Japan, 2‒21‒1 Osawa, Mitaka, Tokyo 181‒8588, Japan}
\affiliation[7]{Department of Earth and Planetary Science, University of Tokyo, 7-3-1 Hongo, Bunkyo-ku, Tokyo 113-0033, Japan}
\affiliation[8]{Aix Marseille Universit\'e, CNRS, CNES, LAM, Marseille, France}
\affiliation[9]{Jet Propulsion Laboratory, California Institute of Technology, Pasadena, CA 91109, USA}
\affiliation[10]{University of Colorado Boulder, Laboratory for Atmospheric and Space Physics, 3665 Discovery Drive, Boulder CO, 80303, USA}
\affiliation[11]{Kapteyn Astronomical Institute, University of Groningen, Landleven 12, 9747\,AD Groningen, Netherlands}
\affiliation[12]{LPC2E, OSUC, Univ Orleans, CNRS, CNES, Observatoire de Paris, F-45071 Orleans, France}
\affiliation[13]{Observatori Astron\`omic, Universitat de Val\`encia, Parc Cient\'ific, C. Catedr\'atico Jos\'e Beltr\'an 2, E-46980 Paterna, Val\`encia, Spain}
\affiliation[14]{Institut de Ci\`encies de I’Espai (ICE-CSIC), Campus UAB, Carrer de Can Magrans s/n, 08193 Cerdanyola del Vall\`es, Catalonia, Spain}
\affiliation[15]{Institut d’Estudis Espacials de Catalunya (IEEC), 08860 Castelldefels, Barcelona, Catalonia, Spain}
\affiliation[16]{Instituto de Astrof\'isica de Andaluc\'ia (IAA-CSIC), Glorieta de la Astronom\'ia s/n, E-18008 Granada, Spain}
\affiliation[17]{Instituto de Estudios Astrof\'isicos, Facultad de Ingenier\'ia y Ciencias, Universidad Diego Portales, Av. Ej\'ercito Libertador 441, Santiago, Chile}
\affiliation[18]{Millennium Nucleus on Young Exoplanets and their Moons (YEMS)}

\abstract{The majority of the Solar System planets are sources of bright radio emission, driven by energetic electrons trapped within each planet's magnetic field. Detection of this emission from exoplanets provides a unique opportunity to characterise their magnetic fields, which is key to determining the atmospheric evolution of exoplanets. However, a conclusive detection of radio emission from an exoplanet remains at large, primarily due to a lack of sensitivity at low radio frequencies. On the other hand, planet-like radio signatures have been detected on objects called ultracool dwarfs (UCDs) for over two decades. UCDs are of comparable sizes to Jupiter, but are more massive. They also possess similar interior structures to Jupiter, the region where magnetic fields are generated. Therefore, UCDs are ideal targets to study to advance our understanding of how magnetic fields manifest at planetary scales. In this Chapter, we outline the revolutionary role that the Square Kilometre Array will play in the study of exoplanets and UCDs. We anticipate that it will facilitate the first detection of radio emission from giant exoplanets with strong magnetic fields, and will deliver thousands of detections of UCDs within a few hundred parsecs. Combined with very long baseline interferometry, we also expect that astrometric monitoring will enable the detection of planets of a few Earth masses orbiting nearby radio-emitting UCDs. These findings will open a new window into how planets form and evolve in extrasolar systems.}

\begin{document}

\maketitle


\section{Introduction}

Detecting and characterising planets in extrasolar systems, known as exoplanets, is key to understanding the diversity of planetary systems and placing our own Solar System in context. The demographics and atmospheric compositions of exoplanets inform us about their formation \citep{fulton18}, evolution \citep{schlecker22, stefansson23}, and potential habitability \citep{espinoza25}. While we have now discovered over 6\,000 exoplanets, current detection methods heavily skew our statistics towards massive planets that orbit close to their host stars in edge-on configurations from our vantage point. Additionally, we currently lack the technology to directly detect the magnetic fields of exoplanets \citep{oklopcic20, benjaffel22}, which is key to regulating planetary atmospheric escape \citep{owen14, dong19, carolan21} and ultimately habitability in the case of terrestrial planets \citep{blackman18}.

Brown dwarfs and late M~dwarfs (effective temperatures $\lesssim2700$~K), collectively known as ultracool dwarfs (UCDs), provide unique insights into the magnetic fields and atmospheres of exoplanets. With sizes comparable to Jupiter but masses about 13 to 70 times greater \citep{davis25}, thousands of UCDs have been discovered in our astronomical backyard\footnote{\href{https://doi.org/10.5281/zenodo.15802304}{The UltracoolSheet: Photometry, Astrometry, Spectroscopy, and Multiplicity for 4000+ Ultracool Dwarfs and Imaged Exoplanets}}. The cool atmospheres of UCDs are excellent analogues for those of gas giant exoplanets, and can be studied in isolation from the irradiation of a host star \citep{vos23, faherty24}. UCD demographics also provide clues about how both stars and planets form \citep{palau24}. They also possess powerful magnetic fields that can be detected with current technology, which allow us to probe their interior structures. Since the structure, composition, and convective properties of UCD interiors are thought to resemble those of Jupiter \citep{showman20}, studying magnetism on UCDs provides insights into the magnetic field generation and interior structures of exoplanets, the latter of which are notoriously difficult to study \citep{vandijk25}.

In this Chapter, we discuss the revolutionary role the Square Kilometre Array (SKA) will play in the detection and study of exoplanets and UCDs. This is anticipated for two reasons. First, bright low-frequency radio emission is associated with all magnetised planets in the Solar System \citep{zarka98}. Second, emission reminiscent of that observed on the Solar System planets is prevalent on UCDs. We expect that the unprecedented sensitivity of the SKA will deliver the first concrete detections of radio emission from magnetised giant planets, and will yield thousands of detections of UCDs within a few hundred parsecs from Earth. Additionally, long-term monitoring with the SKA in combination with very long baseline interferometry will enable astrometric detection of exoplanets orbiting radio-loud stars and UCDs. The wealth of new science enabled by the SKA will cement radio astronomy as a core technique for discovering and characterising exoplanets and UCDs in the coming decades.


\section{Radio emission from planets in the Solar System}

To appreciate the SKA's potential for discovering and characterising exoplanets and UCDs, we first provide an overview of radio emission from the planets in the Solar System. Most Solar System planets with intrinsic magnetic fields (Earth, Jupiter, Saturn, Uranus, Neptune) are sources of bright low-frequency radio emission \citep{zarka98}. This radio emission is particularly prevalent from Jupiter, which outshines the quiet Sun by up to three orders of magnitude at frequencies below 40~MHz \citep{zarka07}. Planetary radio emission is primarily driven by the electron cyclotron maser instability \citep{treumann06}. Jupiter is also a known source of broadband emission from $\sim10$~MHz to 10~GHz, which is generated via synchrotron processes \citep{depater81, girard16}. We discuss the core physics underpinning these processes in Sections~\ref{sec:SS aurorae} and \ref{sec:SS radiation belts}.


\subsection{Auroral emission}
\label{sec:SS aurorae}  

Electron cyclotron maser (ECM) emission is driven by the amplification of electromagnetic waves by semi-relativistic (keV) electrons in strongly magnetised, low density plasmas. First proposed as the driving mechanism of the Earth's kilometric ($\sim$100 to 600~kHz) radio emission \citep{wu79}, ECM has been validated for the Earth, Jupiter, and Saturn from a wealth of in-situ observations \citep[e.g][]{louarn90, lamy10, collet24}. Given its prevalence in the Solar System, ECM is thought to be widely applicable to magnetised exoplanets, as well as UCDs and stars given their similar plasma properties \citep{treumann06, zarka07}.

When electrons are accelerated in a planetary magnetic field, they travel towards the magnetic poles. If an electron's velocity vector is predominantly aligned with the magnetic field, it will precipitate into the atmosphere and collide with atmospheric molecules. These collisions can produce bright emission from infrared to X-ray wavelengths \citep{badman15, saur21}. Electrons that do not precipitate however, i.e. those with velocity vectors that are predominantly misaligned (perpendicular) to the magnetic field, can amplify electromagnetic waves close to the electron cyclotron frequency \citep{treumann06, collet24}:
\begin{equation}
\nu_\text{c} = \frac{eB}{2\pi m_e} \approx2.8 \times \frac{B}{1~\text{Gauss}}~\text{MHz} ,
\label{eq:cyclotron frequency}
\end{equation}
where $e$ and $m_e$ are the electron charge and mass, and $B$ is the local magnetic field strength. On the Earth, the optical signature of the precipitating electrons manifests as the well-known phenomenon of the aurora borealis and australis. Hence, this emission is generally referred to as `auroral' emission, irrespective of wavelength. We adopt this terminology from here onwards. 

Auroral radio emission is highly circularly/elliptically polarised \citep{zarka98, zarka04b, lamy20, zarka21}, and provides a direct way to constrain the field strengths of magnetised bodies (Equation~\ref{eq:cyclotron frequency}). Jupiter exhibits the strongest magnetic field of all the Solar System planets. With a maximum strength of $\sim 14$~G, its auroral emission extends up to $\sim40$~MHz \citep{zarka07}. A condition for radio aurorae to be generated is that the cyclotron frequency at the emission site exceeds the plasma frequency:
\begin{equation}
\nu_\text{p} = \sqrt{\frac{e^2 n_e}{\pi m_e}} \approx 9\times10^{-3} \sqrt{\frac{n_e}{1~\text{cm}^{-3}}}~\text{MHz} ,
\label{eq:plasma frequency}
\end{equation}
where $n_e$ is the free electron density. In other words, detection of auroral emission also provides a constraint on the local electron density. Recent in-situ observations of Jupiter taken by \textit{Juno} indicate that $\nu_\text{c} \gtrsim 3 \nu_\text{p}$ in order for ECM emission to be generated \citep{collet25}.

Another defining characteristic of auroral radio emission is its beam pattern, which is highly anisotropic. Emission is beamed along the surface of a cone that is centered on the magnetic field line at the emission site, with an opening angle $\leq 90^\circ$. This causes auroral radio emission to manifest as short pulses in time as the emission cone rotates in and out of view of the observer. This characteristic can be exploited to infer the geometry of the magnetic field at the source location \citep{lamy08b, kuznetsov12, louis19, bastian22, kavanagh24}.

Not long after \textit{Voyager~1} discovered auroral radio emission on Saturn \citep{kaiser80}, \cite{desch84} reported that the brightness of radio aurorae on the magnetised Solar System planets appear to be strongly correlated with the solar wind energy flux intercepted by each planet's magnetic field. Follow-up studies \citep[e.g.][]{zarka01, zarka07} interpreted this correlation as the intercepted solar wind energy regulating the energy released via magnetic reconnection in the nightside of the planet's magnetic field \citep{dungey61}, providing the source of the keV electrons necessary to power aurorae. By analysing data from all the Solar System planets and moons with intrinsic magnetic fields, \citet{zarka18} demonstrated that any magnetised body's auroral radio power scales linearly with the \textit{magnetic} energy flux of the plasma flow it intercepts. This scaling law is often referred to as the radio-magnetic scaling law.

The magnetic power of the solar wind that is intercepted by a magnetised planet is \citep{zarka18}
\begin{equation}
P_\text{mag} = \frac{{B_\text{w}}^2\sin^2\theta}{\mu_0} \times \Delta u \times \pi {R_\text{m}}^2 ,
\label{eq:magnetic power}
\end{equation}
where $B_\text{w}$ is the magnetic field strength of the solar wind, $\mu_0$ is the vacuum permeability, and $\Delta u$ is the velocity of the solar wind relative to the orbital motion of the planet. $\theta$ is the angle between the direction of $B_\text{w}$ and $\Delta u$, both of which are vectors in three dimensions. $R_\text{m}$ is the size of `magnetosphere' \citep{bagenal13}, the region where the planet's magnetic pressure dominates. This can be estimated by balancing the magnetic pressure of the planet with the total pressure of the incident solar wind $p_\text{w}$. If the planetary magnetic field is dipolar and with its axis perpendicular to the inflowing wind, one can show that \citep{voigt95, griessmeier04}
\begin{equation}
\frac{R_\text{m}}{R_\text{p}} = \Big(\frac{f_0{B_\text{p}}^2}{4\mu_0 p_\text{w}}\Big)^{1/6}
\label{eq:magnetosphere}
\end{equation}
where $R_\text{p}$ is the planetary radius, $B_\text{p}$ is the planetary magnetic field strength at its magnetic poles, and $f_0$ is the magnetopause form factor. For a realistic magnetosphere with a magnetotail, $f_0 = 1.16$ \citep{voigt95}. Equation~\ref{eq:magnetic power} can be thought of as the flow of Poynting flux through the magnetospheric cross-sectional area $\pi {R_\text{m}}^2$ at a flow velocity of $\Delta u$. In the Solar System, $\sim0.2\%$ of the magnetic energy flux intercepted by each planet's magnetosphere is estimated to be converted into radio emission \citep{zarka25b}.

There are also additional drivers of aurorae on some of the magnetised Solar System planets. For instance, Jupiter is known to magnetically interact with its innermost moons through `sub-Alfv\'enic' interactions \citep{zarka07, saur13}, which accelerate electrons from the surrounding plasma environment. These electrons accelerate along the planet-moon flux tube, powering auroral ECM emission as described above. The point where the surrounding plasma ceases to rigidly co-rotate with Jupiter is also thought to be a source of electron acceleration that can drive the generation of aurorae \citep{hill01, cowley01}. We note however that the role this mechanism plays in driving auroral emission has been disputed \citep[see][]{bonfond20, yao22, sicorello25}.


\subsection{Radiation belt emission}
\label{sec:SS radiation belts}

The magnetospheres of the Earth, Jupiter, Saturn, Uranus, Neptune and Ganymede are also known to trap energetic charged particles in regions near their magnetic equators, known as radiation belts \citep{mauk10, kollmann22}. The kinetic energies of these particles are measured to range from $\sim10$~keV to tens of MeV. Electrons in this energy regime can power broadband radio emission via (gyro)synchrotron processes. This emission is brightest close to the critical frequency \citep{longair11}: 
\begin{equation}
\nu_\text{crit} = \frac{3}{2}\gamma^2 \nu_\text{c} \sin\alpha ,
\label{eq:synchrotron critical frequency}
\end{equation}
where $\gamma$ is the Lorentz factor and $\alpha$ is the angle between the magnetic field and the electron's velocity vector.

As synchrotron emission is an incoherent emission process, it does not produce the same brightness levels achieved by the coherent ECM mechanism. Nevertheless, the high energy electrons trapped within Jupiter's magnetosphere produce synchrotron emission that has been detected from $\sim10$~MHz to $10$~GHz \citep{bolton02, girard16}. This emission is at least 5 orders of magnitude dimmer than Jupiter's auroral emission \citep{girard16}, and exhibits significant linear polarisation \citep{bolton02} – a characteristic trait of synchrotron emission \citep{rybicki86}. The emission pattern of synchrotron emission is also highly collimated, which induces rotational modulation in Jupiter's radiation belt emission \citep{velusamy20, matuszewska22}.

While trapped energetic particles have also been measured in-situ for the Earth, Saturn, Uranus, Neptune, and Ganymede \citep{mauk10, kollmann22}, their associated synchrotron emission so far remains undetected. This is likely due to the weaker magnetic fields, which place their emission below the Earth's ionospheric cutoff \citep[Equation~\ref{eq:synchrotron critical frequency};][]{hegedus20, marc24}, and/or lower particle energies. The prominent ring around Saturn is also thought to inhibit the build-up of energetic electrons in its radiation belt \citep{roussos14}. Finally, while Mercury in principle could also have a radiation belt \citep{lukashenko20}, its magnetosphere is very small due to its weak field and proximity to the Sun \citep{bagenal13}. The BepiColombo mission may soon shed new light on this topic \citep[e.g.][]{williamson26}.


\section{Radio emission from exoplanets and ultracool dwarfs}

\begin{figure}
\centering
\includegraphics[width = 0.75\linewidth]{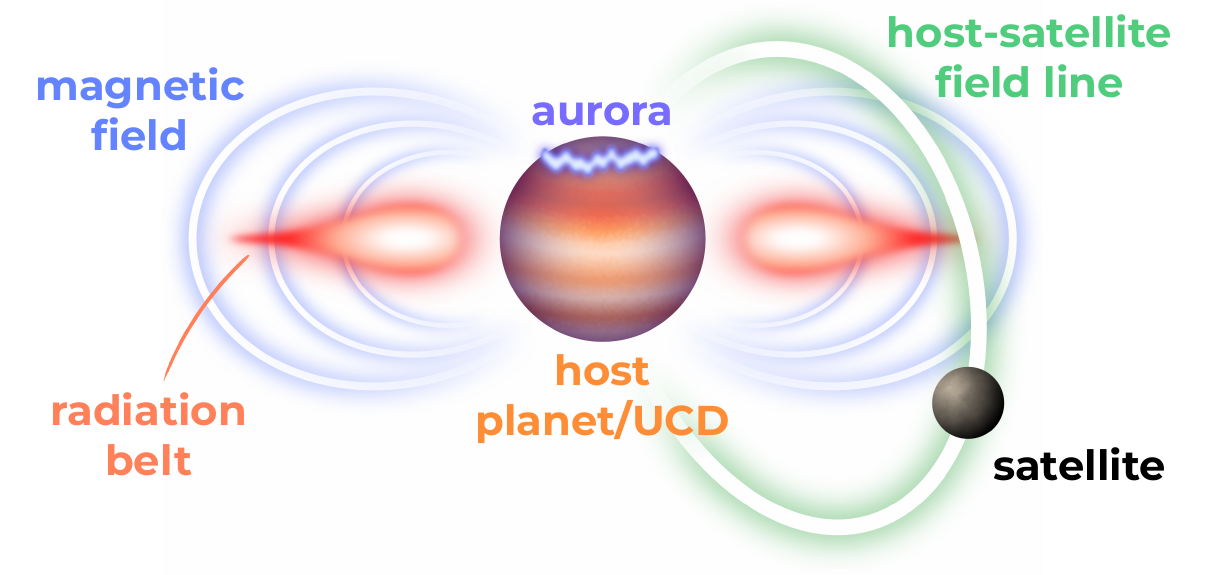}
\caption{Sketch highlighting the different windows for studying and characterising exoplanets and ultracool dwarfs (UCDs) with the SKA. In the center, we see a magnetised planet/UCD. Electrons accelerated within the magnetic field lines can produce bright `auroral' radio emission near the magnetic poles. This auroral emission can facilitate the detection and characterisation of magnetised exoplanets and UCDs. Strong magnetic fields can also trap relativistic electrons in a surrounding radiation belt that produces synchrotron radio emission, offering an additional detection pathway for UCDs. There are a variety of potential sources of the energetic electrons powering the radio emission, such as plasma outflow from a nearby star. Another source could be an orbiting satellite that is outgassing material, which forms an equatorial disk that the host's magnetic field interacts with. Satellites orbiting these objects can also induce astrometric motion in the radio emission of the host, which could be spatially-resolved using very long baseline interferometry. This presents the opportunity to detect unseen satellites. The same technique can also be applied to a radio-emitting host star.}
\label{fig:sketch}
\end{figure}

The prevalence of auroral radio emission from the magnetised Solar System planets naturally suggests that exoplanets could also be radio emitters. Additionally, GHz radio emission reminiscent of that seen on the magnetised Solar System planets has been detected from UCDs for over two decades. In this Section, we give an overview of the expected radio signatures from exoplanets, as well as those observed from UCDs. A graphical overview of the radio emission expected from exoplanets and UCDs is also provided in Figure~\ref{fig:sketch}.

\subsection{Exoplanets}
\label{sec:EP radio emission}

Many of the exoplanets discovered to date are giant planets\footnote{Planets with masses greater than $\sim10$ Earth masses – the estimated upper limit for a planet with a rocky composition \citep{parc24}.} in short orbits around Sun-like stars – main-sequence stars with masses between 0.075 and $\sim1.8$ solar masses \citep{robrade09, chabrier23}. Like our Sun, these stars possess magnetically-heated coronae that drive outflows of material called stellar winds \citep{shoda23}. Giant close-in planets intercept much higher stellar wind fluxes than the Solar System planets. If the established radio-magnetic scaling law holds for these systems, these planets could exhibit bright radio aurorae \citep{zarka97, zarka98, zarka25b}.

The expectation of bright radio aurorae from giant, short-orbit planets has spurred a number of radio campaigns aimed at detecting these signatures \citep[e.g.][]{lazio07, smith09, lazio10, hallinan13, murphy15, lynch18}. Despite these efforts, a conclusive detection remains at large \citep[although see][for tentative results]{turner21, zhang25}. A number of reasons have been proposed to explain this \citep{callingham24}. For instance, the stellar wind conditions of Sun-like stars are notoriously difficult to constrain \citep{wood21}, inhibiting accurate predictions for their expected radio powers via Equation~\ref{eq:magnetic power}. Dense stellar winds and extended planetary atmospheres could also attenuate or even inhibit the generation of auroral emission \citep{vidotto17, daleyyates18, weber18, kavanagh19}. The transient nature of auroral emission \citep{hess11, lamy23} also implies that unfavourable geometry and/or insufficient phase coverage could hinder detection of radio emission from exoplanetary systems, although this is becoming less of an issue with the advent of long-term wide-field radio monitoring \citep[e.g.][]{callingham21, tasse26}. Radio frequency interference also poses a significant challenge in searching for exoplanet aurorae comparable to that of Jupiter \citep{cordun25}.

One of the major reasons inhibiting a conclusive detection of an exoplanet's aurora is arguably the ambiguity about their magnetic field strengths. Magnetic fields are thought to be prevalent on giant exoplanets \citep{brain24}, as they have sufficiently large reservoirs of conductive fluid that can convect and generate the field via dynamo action. While it is thought that the magnetic field strength of any planet scales with its internal heat flux provided it rotates sufficiently fast \citep{christensen09}, estimates for giant planets range from around 0.1 G to 1 kG depending on their age, stellar irradiation, and conductivity \citep{reiners10, yadav17, zaghoo18, kilmetis24}. These quantities are generally poorly constrained or unknown for exoplanets. Additionally, the validity of these estimates have been disputed more recently \citep{kao16, kavanagh24}. These uncertainties mean that the appropriate observing frequency for detecting a given exoplanet's aurora is largely unconstrained (Equation~\ref{eq:cyclotron frequency}).

The existence of magnetic fields on rocky planets is more uncertain, as the conditions to sustain a dynamo strongly depend on the planet's interior structure and composition \citep[e.g.][]{blaske21}. Even if the conditions are suitable, \citet{bonati21} estimated that the magnetic fields of rocky planets are at most 10 times as strong as that of the Earth (around 3~G), placing their auroral emission below 8~MHz. Since the high density of free electrons in the Earth's ionosphere reflects radio waves with frequencies below 10~MHz, detecting aurorae on rocky exoplanets will likely require space-based radio infrastructure \citep[e.g.][]{burns21, knapp24}.

We note that auroral emission from the known exoplanet population could also be driven by co-rotation breakdown or sub-Alfvénic interactions with a satellite. However, both of these mechanisms require the presence of a satellite such as an exomoon, which so far have yet to be detected \citep{heller24, kipping25}. Additionally, these mechanisms are not thought to produce observable fluxes without assuming extreme conditions \citep[e.g.][]{nichols11}. Therefore, we do not consider these scenarios any further. Planets orbiting close to their host stars are also thought to generate aurora-like emission on the star via sub-Alfvénic interactions \citep{saur13}. However, this topic is covered in detail in \citet{Vedantham01.2026.SKA}.

Finally, radiation belt emission is also possible for exoplanets. However, Jupiter's belt emission is over 5 orders of magnitude weaker than its auroral emission. As we will show in Section~\ref{sec:EP aurora yields}, the number of exoplanets with aurorae that are expected to be within the detection capabilities of the SKA is low. Therefore, if these planets have Jupiter-like radiation belts, they are likely going to be very challenging to detect even for the SKA.


\subsection{Ultracool dwarfs}
\label{sec:UCD radio emission}

Unlike exoplanets, radio emission has been detected from UCDs for over two decades \citep{berger01}. They generally show two distinct forms of emission: bright, short-duration, circularly-polarised bursts, and weakly polarised quiescent emission. The consensus is that the bursting emission is driven by auroral processes, whereas the quiescent emission is thought to be synchrotron emission originating from radiation belts \citep{kao23, climent23}. In the absence of conclusive radio detections from exoplanets, UCDs serve as suitable analogues to study magnetic field generation at planetary scales.

So far, the bulk of radio detections of UCDs have come at GHz frequencies \citep{williams18}, although advances at low radio frequencies have begun to discover emission from UCDs in the MHz regime \citep[][]{zic19, vedantham20, vedantham23, yiu25}. Their GHz auroral emission implies that UCDs have magnetic fields that are up to a few kilogauss in strength, orders of magnitude stronger than that of Jupiter (Equation~\ref{eq:cyclotron frequency}). This is thought to be due to their large internal heat fluxes which drive convection and generate their magnetic fields via dynamo action \citep{christensen09, reiners10}. While UCDs are generally expected to harbour dipolar magnetic fields \citep{christensen09}, complex field topologies have also been inferred in some cases \citep[e.g][]{lynch15, kao16}.

Estimates from volume-limited surveys indicate that $\sim5\%$ of all UCDs exhibit auroral radio emission detectable with current radio telescopes \citep{route16b}. Their auroral bursts typically repeat over timescales consistent with their rotation periods inferred from their photometric variability \citep{pineda17}. This is reminiscent of Jupiter's auroral radio emission that is controlled by its interactions with the solar wind, its rotation, and its surrounding plasma disk \citep{marques17, zarka21}. Unlike Jupiter however, auroral UCDs are not found orbiting Sun-like stars, implying that their emission is powered by plasma disk interactions \citep{hallinan15}. The strong dependence of UCD radio aurorae on rotation is also apparent from the onset of their radio detectability at rotation periods less than $\sim3$ hours \citep{pineda17}. Theoretical models also predict that the brightness of aurorae arising from plasma disk interactions increases with the square of the object's rotation rate \citep{nichols11, saur21}. 

The powerful magnetic fields of UCDs also make them effective traps for energetic particles that can form radiation belts. Very recently, \cite{kao23} and \cite{climent23} spatially-resolved the quiescent emission of the UCD LSR~J1835+3259, unveiling a bipolar structure that bears a striking resemblance to Jupiter's radiation belt \citep{bolton02}. This object's auroral emission however remains unresolved, consistent with a compact source size. These results have cemented the paradigm for the Jupiter-like nature of radio emission from UCDs. Follow-up studies aimed at detecting the linearly-polarised component of radiation belt emission from UCDs will help to further solidify this hypothesis.

\begin{figure}
\centering
\includegraphics[width = 1\linewidth]{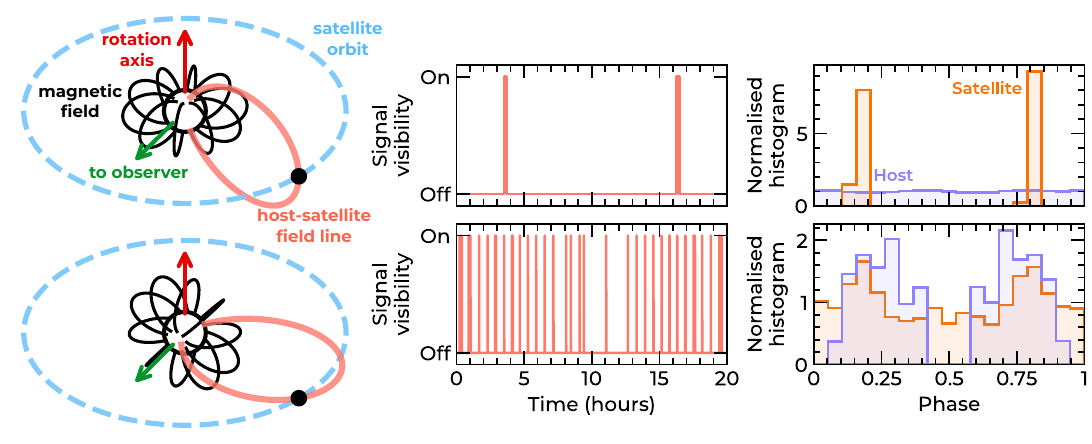}
\caption{The periodicity of satellite-induced radio emission on a magnetised host for different geometric configurations. The top panels correspond to the configuration where the rotation and magnetic axes of the host (in this case, a UCD) are aligned, whereas the bottom panels show the case where they are misaligned by 90 degrees. The left panels show to-scale sketches of each configuration, and the middle panels show the visibility of the satellite-induced signal over time for each configuration, computed using the \texttt{MASER} code \citep{kavanagh23}. The right panels show the normalised histograms of the phases of the satellite's orbit and the rotation of the UCD when the signal is visible. In the aligned configuration, the signal is predominantly modulated with the satellite's orbital period, whereas in the misaligned configuration, the signal is predominantly modulated at the UCD's rotation period. A misaligned configuration could explain the phenomenon of radio aurorae on UCDs being modulated at the UCD's rotation period. The same phenomenon is also applicable to magnetically-interacting stars and planets \citep[see][]{Vedantham01.2026.SKA}. The UCD is 10~Jupiter masses, 1~Jupiter radius, has a rotation rate of 1.28~hours and a dipole field strength of 1~kG, and is seen edge-on. The satellite's orbit is aligned with the UCD's rotation axis, and has an orbital period of 20~hours. The observing frequency is 1~GHz, and the ECM emission cone has an opening angle of $75^\circ$ and a thickness of $2^\circ$. The histograms are computed for an observing window of 1\,000 hours with 100\,000 points.}
\label{fig:periodicity}
\end{figure}

Despite the prevalence of radio emission from UCDs, the plasma source necessary to drive their emission is a mystery. The Jupiter-like nature of their emission naturally suggests that radio-emitting UCDs are orbited by outgassing satellites. This is supported by planet formation theory, which predicts that UCDs are orbited by rocky planets \citep{liu20}. Additionally, rocky close-in planets orbiting UCDs are expected to undergo outgassing due to tidal forces \citep{barr18} and/or induction heating in the presence of the strong magnetic fields of UCDs \citep{kislyakova23}. If satellites do orbit radio-emitting UCDs, the auroral signature is expected to also be modulated at the satellite's orbital period, as is seen for the Io-induced radio aurora on Jupiter \citep{marques17, kavanagh23}. However, radio aurorae on UCDs so far have only been found to be rotationally modulated. This does not rule out a satellite-driven origin explicitly. For instance, the plasma disk interaction could outshine the satellite-induced signal. Alternatively, certain geometric configurations can produce satellite-induced signals that are modulated at the rotation rate of the UCD and not the satellite's period (see Figure~\ref{fig:periodicity}).

Other scenarios have also been inferred to explain the radio signatures from UCDs. At younger ages and higher masses, it becomes difficult to determine whether a UCD is a brown dwarf or a star. In the latter case, the UCD may still exhibit coronal activity \citep{paudel20, magaudda24}, which could result in stochastic flaring or even a quiescent stellar wind outflow. Both scenarios could load the UCD's magnetosphere with the plasma necessitated by their radio emission \citep{leto21}. However, this picture is difficult to reconcile for objects with spectral types later than mid L. Alternatively, lightning in the atmospheres of UCDs may liberate electrons that are supplied to the magnetosphere \citep{helling11}, although whether such events could supply the number of electrons necessary to sustain the observed radio emission from UCDs is unclear.

In Table~\ref{table:UCD radio detections}, we list the relevant properties of each individual UCD with detected radio emission. Objects in binaries/multiple systems are only listed if their radio emission has been associated with a specific component. There are numerous examples of radio detections from systems containing UCDs where the emitting component has not been identified \citep{kao25}. However, it is not clear in these cases which component emits or if multiple components contribute by different amounts. Therefore, we choose to exclude them for our purposes. The frequency and flux densities listed are indicative values, and can vary in time and frequency.

\begin{table*}
\setlength{\tabcolsep}{2.5pt}
\renewcommand{\arraystretch}{1.03}
\centering
\small
\caption{The catalogue of individual radio-detected UCDs listed in order of discovery. The columns list each UCD's name, spectral type (SpT), distance ($d$), rotation period inferred from its photometric variability ($P_\text{rot}^\text{phot}$) and radio bursts ($P_\text{rot}^\text{radio}$) rounded to two decimal places (if available), detected quiescent flux density ($F_\text{Q}$) and frequency ($\nu_\text{Q}$), and burst flux density ($F_\text{B}$) and frequency ($\nu_\text{B}$). The last column lists the reference for the discovery paper, followed by the references for SpT, $d$, $P_\text{rot}^\text{phot}$, $P_\text{rot}^\text{radio}$, $F_\text{Q}$ \& $\nu_\text{Q}$, and $F_\text{B}$ \& $\nu_\text{B}$. The values in brackets following the rotation periods are the errors on the least significant digit(s). Of the 24 objects listed, 21 exhibit quiescent emission, and 17 exhibit burst emission. Names superscripted with a \dag~have been shortened, with the full names listed at the end of the Table. Entries denoted with ``–'' have no reported values.}
\label{table:UCD radio detections}
\begin{tabular}{lccccccccc}
Name & SpT & $d$ & $P_\text{rot}^\text{phot}$ & $P_\text{rot}^\text{radio}$ & $F_\text{Q}$ & $\nu_\text{Q}$ & $F_\text{B}$ & $\nu_\text{B}$ & References \\
& & (pc) & (hr) & (hr) & (µJy) & (GHz) & (µJy) & (GHz) \\
\hline
LP 944-20 & M9 & 6.427 & 3.84 & – & 100 & 8.5 & 2600 & 8.5 & 1, 2, 3, 4, –, 1, 1 \\
TVLM 513$^\dag$ & M8.5 & 10.734 & 1.96 & 1.96 & 464 & 8.44 & 5000 & 8.44 & 5, 2, 3, 4, 6, 6, 6 \\
J0036+18$^\dag$ & L3.5 & 8.736 & 2.7(3) & 3.08(5) & 240 & 8.46 & 500 & 4.88 & 5, 2, 3, 7, 8, 5, 8 \\
BRI 0021$^\dag$ & M9.5 & 12.446 & 3.05 & – & 25 & 8.46 & 360 & 8.46 & 5, 2, 3, 9, –, 5, 5 \\
J1048-39$^\dag$ & M8 & 4.045 & – & – & 140 & 4.8 & 29600 & 8.64 & 10, 2, 3, –, –, 10, 10 \\
LHS 3003 & M7 & 7.053 & – & – & 270 & 4.8 & – & – & 10, 2, 3, –, –, 10, – \\
J1835+32$^\dag$ & M8.5 & 5.689 & 2.84 & 2.84(1) & 722 & 8.44 & 2500 & 8.44 & 11, 2, 3, 4, 8, 8, 8 \\
J0523-14$^\dag$ & L2.5 & 12.734 & – & – & 231 & 8.46 & – & – & 11, 2, 3, –, –, 11, – \\
LP 349-25 A & M8 & 14.128 & – & – & 200 & 5 & – & – & 12, 13, 3, –, –, 13, – \\
LP 349-25 B & M9 & 14.128 & – & – & 110 & 5 & – & – & 12, 13, 3, –, –, 13, – \\
J0746+20 A$^\dag$ & L0 & 12.352 & 3.32(15) & 3.2(2) & 71 & 5 & 500 & 5 & 14, 15, 15, 16, 15, 15, 15 \\
J0746+20 B$^\dag$ & L1.5 & 12.352 & 2.07 & 2.1(1) & 138 & 5 & 1000 & 5 & 14, 15, 15, 17, 15, 15, 15 \\
J1314+13 B$^{\dag,a}$ & M7 & 16.39 & – & – & 1000 & 5 & 7000 & 5 & 18, 2, 19, –, –, 19, 19 \\
J1047+21$^\dag$ & T6.5 & 10.56 & 1.74(1) & 1.76(1) & 9 & 6 & 1550 & 5 & 20, 2, 21, 22, 23, 23, 23 \\
J1906+40$^\dag$ & L1 & 16.759 & 8.9 & – & 16.5 & 5 & – & – & 24, 2, 3, 24, –, 24, – \\
J0136+09$^\dag$ & T2.5 & 6.118 & 2.39 & $2.88^{+0.34}_{-0.27}$ & 11.2 & 10 & 51.5 & 10 & 21, 2, 3, 26, 27, 27, 27 \\
J1043+22$^\dag$ & L8 & 16.4 & – & $2.21^{+0.14}_{-0.13}$ & 16.3 & 6 & 69.0 & 6 & 21, 2, 21, –, 27, 21, 21 \\
J1237+65$^\dag$ & T6.5 & 10.42 & – & $2.28^{+0.10}_{-0.09}$ & 27.8 & 10 & 160 & 10 & 21, 2, 21, –, 27, 27, 27 \\
J0607+24$^\dag$ & L8 & 7.243 & – & – & 15.6 & 6.05 & – & – & 25, 2, 3, –, –, 25, – \\
J1122+25$^\dag$ & T6 & 17.7 & – & 1.93 & – & – & 2000 & 5 & 28, 2, 29, –, 29, –, 28 \\
J1750-00$^\dag$ & L4.5 & 9.210 & – & – & 56.4 & 6 & 667 & 6 & 30, 30, 3, –, –, 30, 30 \\
Elegast$^{\dag,b}$ & T6.5 & 65 & – & – & 1100 & 0.144 & – & – & 31, 31, 31, –, –, 31, 31 \\
J0623-04$^\dag$ & T8 & 11.44 & – & 1.91(1) & – & – & 3000 & 2.1 & 32, 32, 32, –, 32, –, 33 \\
J0752+16$^\dag$ & M7 & 18.881 & 21.21(5) & – & – & – & 130 & 9 & 34, 34, 3, 34, –, –, 34 \\
\hline
\end{tabular}
\caption*{\scriptsize\textit{References} – 1: \citet{berger01}, 2: \citet{pineda17}, 3: \citet{gaiadr3}, 4: \citet{milespaez23}, 5: \citet{berger02}, 6: \citet{hallinan07}, 7: \citet{metchev15}, 8: \citet{hallinan08}, 9: \citet{dulaimi23}, 10: \citet{burgasser05}, 11: \citet{berger06}, 12: \citet{phanbao07}, 13: \citet{curiel24}, 14: \citet{antonova08}, 15: \citet{zhang20}, 16: \citet{harding13}, 17: \citet{berger09}, 18: \citet{mclean11}, 19: \citet{williams15a}, 20: \citet{route12}, 21: \citet{kao16}, 22: \citet{allers20}, 23: \citet{williams15b}, 24: \citet{gizis13}, 25: \citet{gizis16}, 26: \citet{artigau09}, 27: \citet{kao18}, 28: \citet{route16a}, 29: \citet{williams17}, 30: \citet{richeyyowell20}, 31: \citet{vedantham20}, 32: \citet{rose23}, 33: \citet{kavanagh24}, 34: \citet{magaudda24} \\
$^\dag$ \textit{Full names} – TVLM~513-46546, LSPM~J0036+1821, BRI~0021-0214, DENIS~J104814.6-395606, LSR~J1835+3259, 2MASS~J05233822-1403022, 2MASS~J07464256+2000321~A, 2MASS~J07464256+2000321~B, 2MASS~J13142039+1320011~B, 2MASS~J10475385+2124234, WISEP~J190648.47+401106.8, SIMP~J013656.5+093347.3, 2MASSI~J1043075+222523, 2MASS~J12373919+6526148, WISEP~J060738.65+242953.4, WISEP~J112254.73+255021.5, 2MASS~J17502484-0016151, BDR~J1750+3809, WISE~J062309.94-045624.6, 2MASS~J07522390+1612157 \\
\textit{Other notes} – $a$: The quiescent and bursting values quoted for J1314+13~B come from \citet{williams15a}, who do not spatially-resolve the binary. However, \citet{forbrich16} later showed that component B is the only detectable radio emitter in the system, so we assume the emission reported by \citet{williams15a} also originated from component B. $b$: While this UCD does not show bursting emission, its quiescent emission is fully circularly polarised, a characteristic generally associated with auroral emission processes \citep[see][]{vedantham20}.}
\end{table*}


\section{Detecting exoplanet aurorae with the SKA}
\label{sec:EP aurorae}

As discussed in Section~\ref{sec:EP radio emission}, radio emission from exoplanets that is detectable with the SKA is likely limited to auroral emission from magnetised giant planets, with the most plausible driver of their emission being arguably the winds of their host stars. The state-of-the-art for predicting the strength and frequency of this emission involves solving the magnetohydrodynamics equations describing the stellar wind outflow in three dimensions \citep[e.g.][]{vidotto12, kavanagh19, elekes23}. However, these models are computationally expensive, and the information required to construct them (such as the surface magnetic field topology of the host star) is generally only available in a handful of cases. Lacking this information, simpler stellar models such as the isothermal model developed by \citet{parker58} can provide sufficiently accurate results. This model assumes that the plasma temperature is constant from the stellar corona outwards. The benefit of the isothermal model is that it only depends on two parameters, and can be computed rapidly for a large set of parameters. In this Section, we use this model to predict the emission strength and frequency of all known giant exoplanets and assess their detectability with the SKA.


\subsection{Stellar wind prescriptions for known exoplanet hosts}
\label{sec:EP aurora modelling}

To predict the wind-driven auroral emission of each giant exoplanet, we take a similar approach to  \citet{griessmeier07} and \citet{mauduit23}. We first use the code developed by \citet{kavanagh20} to obtain the stellar wind velocity profile for each system as described by \citet{parker58}. For a given stellar mass and coronal (wind) temperature ($T$), we obtain the wind velocity $u_\text{w}$ as a function of distance from the star $r$. We then compute the mass density profile of the wind via the conservation of mass:
\begin{equation}
\rho_\text{w} = \frac{\dot{M}}{4\pi r^2 u_\text{w}} ,
\label{eq:wind density}
\end{equation}
where $\dot{M}$ is the stellar wind mass-loss rate. With this, we compute the pressure exerted by the stellar wind on each exoplanet:
\begin{equation}
p_\text{w} = \rho_\text{w}(a^2 + \Delta u^2) + \frac{{B_\text{w}}^2}{2\mu_0}.
\label{eq:wind pressure}
\end{equation}
Here $a$ is the sound speed, which depends on the temperature $T$:
\begin{equation}
a = \sqrt{\frac{2k_\text{B}T}{m_\text{p}}} ,
\end{equation}
where $k_\text{B}$ is Boltzmann's constant and $m_\text{p}$ is the proton mass. 

For the magnetic field strength of the wind $B_\text{w}$, we follow the same approach as described in Section~2.2 of \citet{Vedantham01.2026.SKA}. Within the region where the wind's magnetic pressure dominates, i.e.
\begin{equation}
\frac{{B_\text{w}}^2}{2\mu_0} > \rho_\text{w} {u_\text{w}}^2 ,
\label{eq:mag dom}
\end{equation}
we describe the magnetic field as a closed dipole, whose axis is aligned with both the stellar rotation and planetary orbital axes:
\begin{equation}
B_\text{w} = \frac{B_\star}{2} \Big(\frac{R_\star}{r_\text{orb}}\Big)^3 .
\label{eq:closed field}
\end{equation}
Here, $B_\star$ is the dipole field strength of the star at its magnetic poles, $R_\star$ is the stellar radius, and $r_\text{orb}$ is the planet's orbital distance. In this configuration, $B_\text{w}$ points in the meridional direction. Then outside the magnetically-dominated region, we assume the wind magnetic field is purely radial:
\begin{equation}
B_\text{w} = \frac{B_\star}{2} \Big(\frac{R_\star}{r_\text{orb}}\Big)^2 .
\label{eq:open field}
\end{equation}

With the wind velocity and magnetic field profiles obtained for each exoplanet, we then compute the power dissipated onto the planetary magnetosphere via Equations~\ref{eq:magnetic power} and \ref{eq:magnetosphere}. The isothermal wind model used assumes that the wind flows radially outward, perpendicular to the planet's orbital direction. This gives $\Delta u = \sqrt{{u_\text{w}}^2 + {u_\text{p}}^2}$, where $u_\text{p}$ is the planet's orbital velocity. If the planet is in the closed-field region, the angle $\theta$ between $\Delta u$ and $B_\text{w}$ is $90^\circ$. In the open-field region on the other hand, $u_\text{w}$ is parallel to $B_\text{w}$, so $\theta$ is the angle between $u_\text{w}$ and $\Delta u$. To determine if the wind magnetic field is open or closed at the planet's position for a given $B_\star$, we check if Equation~\ref{eq:mag dom} is true assuming a closed field configuration, since the open field strength drops off more slowly as a function of distance than the closed field. Note that we do not account for the rotation of the star here, which will induce an azimuthal component to both the wind velocity and magnetic field \citep[e.g.][]{weber67} and alter the angle $\theta$.

We assume that some fraction $\varepsilon$ of the magnetic power intercepted by each planet's magnetosphere is dissipated over a bandwidth $\Delta\nu$, producing radio emission that is beamed outwards with a solid angle $\Omega$ and received at a distance $d$ \citep{griessmeier07}:
\begin{equation}
F = \frac{\varepsilon P_\text{mag}}{d^2\Omega\Delta\nu} .
\label{eq:flux density}
\end{equation}
If the emission bandwidth $\Delta \nu = \nu_\text{max} - \nu_\text{min}$ is sufficiently wide ($\nu_\text{max} \gg \nu_\text{min}$), then $\Delta \nu \approx \nu_\text{max}$. The maximum emission frequency corresponds to the cyclotron frequency where a dipolar field line of size $R_\text{m}$ meets the surface \citep{vidotto17}:
\begin{equation}
\Delta \nu \approx \nu_\text{max} = 2.8 \Big(\frac{B_\text{p}}{1~\text{Gauss}}\Big) \Big(1 - \frac{3R_\text{p}}{4R_\text{m}}\Big)^{1/2}~\text{MHz}.
\end{equation}
If $R_\text{m}$ is sufficiently large, $\Delta\nu \propto B_\text{p}$, giving the following scaling relation for the auroral radio flux density:
\begin{equation}
F \propto \Big(\frac{R_\text{p}}{d}\Big)^2 {B_\text{p}}^{-1/3} {B_\text{w}}^2 {\Delta u}^2 {p_\text{w}}^{-1/3} .
\label{eq:flux scaling}
\end{equation}


\subsection{Expected detection yields for aurorae on giant exoplanets}
\label{sec:EP aurora yields}

We now compute the expected flux density and frequency of aurorae of all known giant exoplanets as described in Section~\ref{sec:EP aurora modelling}. We obtain the value of each planet's orbital period ($P_\text{orb}$), radius ($R_\text{p}$), and (minimum) mass ($M_\text{p}$), as well as the stellar mass ($M_\star$) and radius ($R_\star$) and the distance ($d$) to the system from the Default Parameter Set of the NASA Exoplanet Archive\footnote{\href{https://exoplanetarchive.ipac.caltech.edu/}{NASA Exoplanet Archive} (accessed on 10 Oct 2025)}. We compute the planet's orbital distance using the orbital period and stellar mass via Kepler's third law, neglecting the planet's eccentricity.

We filter out systems where the values for $P_\text{orb}$, $M_\star$, $R_\star$ or $d$ are missing. If a planet's radius is not known, we compute it via their (minimum) mass (if available) as described in \citet{parc24}. In the cases where the mass is a minimum mass, then the radii computed using it is also a lower limit. We then discard those which are not giant planets (those with masses below 10 Earth masses). We also limit our calculations to systems where the host star is Sun-like, as they are both the most common exoplanet hosts and have thermal winds as described by the model deployed from \citet{parker58} (Section~\ref{sec:EP radio emission}). We note however that evolved stars with dense stellar winds could power aurorae on exoplanets that are much brighter than those possible for main sequence stars \citep{fujii16}. To determine if the host star is `Sun-like', we compare the physical properties of each star to those estimated from main sequence relations derived as functions of stellar mass. From 0.076 to 0.616 solar masses, we use the absolute visual magnitude relation given in Equation~10 of \citet{benedict16}, and above 0.616 solar masses we use the stellar radius and effective temperature relations given in Equations~3 and 4 of \citet{eker24}. We consider the star to be on the main sequence if it is within one standard deviation of these relations. We also only consider systems where the stellar effective temperature is less than 8\,000 K. Above this temperature, main-sequence stars no longer possess magnetically-heated coronae \citep{robrade09}, which are necessary to drive hot magnetised stellar wind outflows akin to that of the Sun. Finally, we discard systems with declinations outside the observing capabilities of the SKA ($>+30^\circ$). These filtering steps bring the number of exoplanets from 6\,028 down to 613.

For each exoplanet, we have four unknowns: the magnetic field strength of the host star $B_\star$, the wind temperature $T$ and mass-loss rate $\dot{M}$, and the planet's field strength $B_\text{p}$. As these parameters are all largely unconstrained, we explore two scenarios each for both the star and the planet. For each planet, we consider both a strong ($B_\text{p} = 100$~G) and weak ($B_\text{p} = 25$~G) field, corresponding to emission frequencies in the observing range of SKA-Low (Equation~\ref{eq:cyclotron frequency}). For each star on the other hand, we consider an `active' and `inactive' state. For the active state, we adopt a strong surface magnetic field of 100~G and a wind mass-loss rate of $1000~\dot{M}_\odot$, where $\dot{M}_\odot = 1.26\times10^{12}$~g~s$^{-1}$ is the solar wind mass-loss rate \citep{cohen11}. The stellar magnetic field strength adopted here is representative of the dipolar magnetic field of an active planet-hosting star \citep[e.g.][]{klein21}, and the mass-loss rate is comparable to the highest values inferred for Sun-like stars to date \citep{jardine19}. For the wind temperature, we use the maximum physical coronal temperature that gives a supersonic outflow at the stellar surface, which is necessary to drive the outflow \citep{kavanagh20}:
\begin{equation}
T < \frac{G M_\star m_\text{p}}{4 k_\text{B} R_\star} . 
\end{equation}
For the systems we consider, this maximum temperature ranges from $\sim4$ to 8 million Kelvin. The high temperature and large mass-loss rate assumed for the active state result in a fast and dense wind for each star, as are inferred for active low-mass main-sequence stars \citep{kavanagh21}. For the inactive state, we adopt a stellar magnetic field of 1~G, a mass-loss rate of $0.1~\dot{M}_\odot$, and a coronal temperature of 0.5~MK.

We now compute the flux density and peak frequency of the aurorae on each giant planet as described in Section~\ref{sec:EP aurora modelling} for each combination of an active/inactive host star and strong/weak planetary magnetic field. In Equation~\ref{eq:flux density}, we assume a magnetic energy conversion efficiency of $\varepsilon = 2\times10^{-3}$ and a solid angle of $\Omega = 1.6$~sr, which are the inferred values from Solar System observations \citep{zarka04a, zarka07}. Given the existing uncertainties on the properties considered here, we do not propagate the errors in the values taken from the NASA Exoplanet Archive.

\begin{figure}
\centering
\includegraphics[width = 0.7\linewidth]{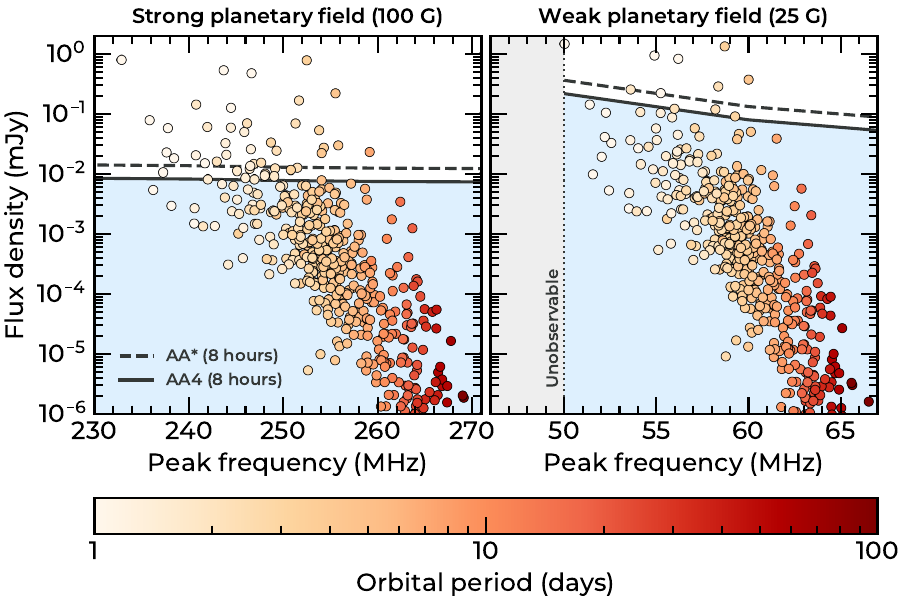}
\caption{Predicted auroral radio signatures of the known giant planet population assuming their host stars are active, for a planetary magnetic field strength of 100~G (left) and 25~G (right). Each planet is coloured based on its orbital period. The detection thresholds of SKA-Low in the AA* and AA4 configurations for an 8-hour image exposure are also shown as dashed and solid lines.}
\label{fig:EP aurorae}
\end{figure}

In Figure~\ref{fig:EP aurorae}, we show the expected flux density and peak frequency of the auroral emission on each giant exoplanet for a strong and weak planetary field. We only show the results assuming an active host star, as all systems for an inactive state are below the detection threshold of SKA-Low in configuration AA4 for an 8-hour image observation. While weaker planetary magnetic fields produce slightly higher fluxes (see Equation~\ref{eq:flux scaling}), the sensitivity of SKA-Low is significantly lower at the corresponding peak frequencies ($\sim60$~MHz). Additionally, conditions are more favourable for generating ECM emission if the field strength is higher (Section~\ref{sec:SS aurorae}). We also note that absorption of the auroral emission by the wind of the host star is likely to be more severe at lower frequencies \citep{vidotto17, kavanagh19}. However, emission from planets around active stars may be absorbed substantially more than those around inactive stars, as active stars likely possess dense stellar winds \citep[e.g.][]{jardine19}.

We also estimate the number of systems detectable with SKA-Low for different array configurations and observing times. For this, we compute the image noise of the SKA as outlined in Section~8 of \citet{braun19}, linearly interpolating the array sensitivities for each frequency\footnote{The most recent estimates for the sensitivities of SKA-Low and Mid are given in Tables 5 and 6 of the Anticipated SKA1 Science Performance document (\href{https://www.skao.int/sites/default/files/documents/SKAO-TEL-0000818-V2_SKA1_Science_Performance.pdf}{Revision 002}).}. We consider a system to be detected if it reaches a signal to noise ratio (SNR) of 5 or greater. In Table~\ref{table:EP aurora yields} we list the number of systems detectable with SKA-Low in array configurations AA* and AA4, for observing times of 1 and 8~hours. Overall, prospects are more favourable in the strong field scenario and longer integration times. With weak planetary magnetic fields, the only planet detectable is $\tau$~Boo~b, requiring at least an 8 hour integration for a conclusive detection. With strong planetary magnetic fields however (corresponding to higher frequencies, and consequently greater sensitivity), more planets become detectable. This implies that SKA-Mid could be better suited for detecting exoplanets, although the field strengths detectable with SKA-Mid may only be attainable for young, massive Jupiter-like planets in wide ($>1$~au) orbits \citep{kilmetis24}. Irrespective of the observing time and array configuration, the five brightest planets in the strong field regime are $\tau$~Boo~b, TOI-2109~b, HIP~65~A~b, WASP-18~b, and 51~Peg~b. We note that the auroral signatures modelled here are likely short in duration due to the intrinsic beaming of the emission mechanism \citep{ashtari22}, which we do not account for. As a result, a single integration may smear out the signal. Therefore, conclusive detection may require stacking multiple observations of a given system.

While the prospects presented here for detecting exoplanet aurorae with the SKA are uncertain, wide field surveys at low frequencies will allow us to assess the uncertainties on the stellar wind, planetary magnetic field, conversion efficiency, and beaming geometry of each known exoplanet system at an unprecedented level. These surveys may also lead to the detection of aurorae from undetected planets, such as those with long periods in face-on orbits. Due to the uncertainties on the relevant physical parameters and beamed nature of the emission, blind surveys may be the most promising pathway to discovering exoplanet aurorae \citep[e.g.][]{callingham21, tasse26}.

\begin{table}[]
\centering
\caption{The number of known giant exoplanets detectable with the SKA at an SNR > 5. Results are shown for the AA* and AA4 configurations of the SKA for different observing times and planetary field strengths. Note that these numbers are computed under the assumption that the host star of each planet is in an `active' state, with a strong surface magnetic field and dense, fast stellar wind (see text).}
\label{table:EP aurora yields}
\begin{tabular}{ccccc}
Planet field strength & 1 hour (AA*) & 8 hours (AA*) & 1 hour (AA4) & 8 hours (AA4) \\
\hline
Weak (25~G) & 0 & 1 & 0 & 3 \\
Strong (100~G) & 5 & 11 & 8 & 16 \\
\end{tabular}
\end{table}


\section{Detecting ultracool dwarf aurorae with the SKA}
\label{sec:UCD aurorae}

The observed radio aurorae on UCDs (Table~\ref{table:UCD radio detections}) can inform us about their expected detection yields with the SKA, without utilising scaling laws that contain unknown parameters as in Section~\ref{sec:EP aurorae}. With sufficient telescope sensitivity, the detectability of auroral emission is limited by the viewing and magnetic geometry \citep{hess08, louis19, kavanagh23, lamy23}. On UCDs, auroral emission is generally interpreted as originating from either a shell of field lines \citep{leto16, pineda17, bloot24, guirado25} or individual `active field lines' (AFLs) within a dipolar \citep{kuznetsov12, bastian22, kavanagh24} or multipolar field geometry \citep{lynch15}. Of all these different scenarios, the shortest visibility window for the emission is that on a single AFL. A number of numerical tools have been developed to compute the visibility of auroral emission from magnetised bodies like UCDs, such as \texttt{ExPRES} \citep{hess08, louis19} and \texttt{MASER} \citep{kavanagh23}. The \texttt{CHARM}\footnote{\url{https://github.com/robkavanagh/charm}} framework recently developed by \citet{kavanagh24} contains different physical models for the origins of auroral emission within dipolar magnetic fields.  \texttt{CHARM} is designed to efficiently explore large parameter spaces, making it well-suited for parametric inversion studies and Monte Carlo simulations. One such model included predicts emission from AFLs. In this Section, we use this model to estimate the detectability of the population of UCDs with the SKA, assuming they all exhibit radio aurorae.


\subsection{Simulating a population of auroral UCDs}
\label{sec:UCD aurora sim}

To estimate the number of UCD aurorae detectable with the SKA, we use the AFL model from \texttt{CHARM} to simulate the radio lightcurves of a population of UCDs within 300~pc, comparable to the scale height of the Milky Way \citep{aganze22}. There are a number of parameters required to use this tool. We describe our sampling method for each one below:

\begin{itemize}[leftmargin = *]
    
    \item \textit{Viewing angle}: The viewing angle or inclination of the system is defined by the angle $i$ formed between our line of sight and the UCD's rotation axis. To randomly orient the rotation axis of each UCD, we uniformly sample $\cos i$ \citep{kavanagh23}.

    \item \textit{Magnetic obliquity}: This describes the angle $\beta$ between the rotation and dipole axis of the UCD. While there may be an underlying distribution of magnetic obliquities relating to the dynamo processes on these objects, it is unknown. Therefore, we opt to uniformly sample $\beta$ from 0 to 90$^\circ$.

    \item \textit{Field strength}: Like the obliquity, there is presumably some underlying distribution of field strengths for UCDs, which is unknown. So far, UCD aurorae have been predominantly detected from $\sim1$ to 20~GHz \citep{kao18, rose23}, corresponding to dipolar field strengths ($B_0$) of $\sim350$~G to a few kG. To keep our estimates consistent with this range, we draw samples for $B_0$ from a log uniform prior ranging from 350~G to 10~kG. We note however that \citet{vedantham23} recently reported detection of auroral bursts from a binary UCD system at 144~MHz, suggesting that some UCDs may have magnetic fields much weaker than our adopted range.

    \item \textit{Rotation period}: The rotation periods of UCDs with spectral types from mid-L to late-T are estimated to follow a log-normal distribution \citep{radigan14}:
    \begin{equation}
    f(P_\text{rot}) = \frac{1}{P_\text{rot}\sigma\sqrt{2\pi}} \exp \bigg\{ -\frac{1}{2} \Big(\frac{\ln P_\text{rot} - \mu}{\sigma}\Big)^2 \bigg\} ,  
    \end{equation}
    where $P_\text{rot}$ is in hours, $\sigma = 0.48$ and $\mu = 1.44$. While \citet{route17} have extended this distribution to late-M UCDs, this can increase the risk of including UCDs that are above the hydrogen-burning limit. These UCDs likely possess coronae that drive stellar winds, which will spin down the UCD over time \citep[e.g.][]{popinchalk21}. For instance, the most recently-detected radio-emitting UCD (the M7 J0752+16) has an estimated rotation period of 21~hours (Table~\ref{table:UCD radio detections}). According to the distribution above, the probability of a UCD possessing a rotation period exceeding 20 hours is 0.05\%. Additionally, an M7 UCD must be younger than $\sim100$~Myr in order to be below the hydrogen burning limit, whereas mid-Ls can have ages up to $\sim1$~Gyr \citep{davis25}. J0752+16 also does not show any lithium absorption \citep{wang22}, indicative of a stellar nature. Therefore, we draw the values of $P_\text{rot}$ for each UCD from the distribution presented by \citet{radigan14}. We also note that a period of less than 1 hour is below the breakup limit for objects with masses less than 10 Jupiter masses. 

    \item \textit{Rotation phase}: At the start of each observation, we randomly choose the initial rotation phase of the UCD $\phi_0$ from 0 to $2\pi$. The rotation phase is then:
    \begin{equation}
    \phi_\text{rot} = \phi_0 + \frac{2\pi t}{P_\text{rot}}
    \end{equation}
    When $\beta \neq 0^\circ$, the magnetic axis precesses about the rotation axis, with its position described by $\phi_\text{rot}$ \citep{kavanagh24}.

    \item \textit{Active field line properties}: The AFL is defined by its magnetic longitude $\phi_\text{B}$ and spatial extent or `loop size' $L_\text{B}$ in the magnetic equator \citep{kavanagh24}. For the longitude, we uniformly draw values from 0 to $2\pi$, whereas for the loop size, we choose values from 1 to 100 times the radius. The physical meaning of the loop size relates to the region in the magnetosphere where the electrons are accelerated \citep[e.g.][]{zarka25a}. We currently do not have any means to place constraints on these values. This likely requires solving the magnetohydrodynamic equations describing the flow of plasma in the UCD's magnetosphere for different rotation rates and magnetic geometries.

    \item \textit{Emission cone properties}: An ECM emission cone at a given frequency is defined by its opening angle $\alpha$ and thickness $\Delta\alpha$ \citep{louis19}. We opt for fixed values of $\alpha = 75^\circ$ and $\Delta\alpha=1^\circ$, in line with those inferred for Jupiter \citep{hess08, lamy21}.

\end{itemize}

We also require the following parameters to estimate if the system is visible:

\begin{itemize}[leftmargin = *]
    
    \item \textit{Distance}: We draw the distance to each UCD from 2 to 300~pc assuming a constant space density. We achieve this by drawing the volume $V$ of the spherical shell formed between 2~pc and $d$ from a uniform prior. The lower limit adopted for the distance is that of binary system Luhman~16, the closest known UCDs \citep{luhman13}.

    \item \textit{Luminosity}: The radio luminosity $l$ of each UCD depends on its observed flux densities $F$ and distance $d$:
    \begin{equation}
    l = F d^2 ,
    \label{eq:flux luminosity}
    \end{equation}
    which ranges from $\sim2\times10^{11}$ to $2\times10^{14}$~erg~s$^{-1}$~Hz$^{-1}$ for the UCDs listed in Table~\ref{table:UCD radio detections}. For each simulated UCD, we sample $l$ from a log uniform distribution in this range, and then compute the received flux density by dividing by its sampled distance squared. This implicitly assumes that the solid angle of the emission is in the same range as for the detected UCDs, which should relate to the emission cone properties \citep[see Equation 7 of][]{kavanagh22}.

    \item \textit{Observing frequency}: We uniformly sample the observing frequency $\nu_\text{obs}$ from 1 to 1.76 and 4.6 to 15.3~GHz, corresponding to the frequencies where UCD aurorae have been predominantly detected so far that are observable with the planned configuration of SKA-Mid. The observing frequency maps to the magnetic co-latitude of the emission cone along the AFL, which we compute as described in \citet{kavanagh23}.
    
\end{itemize}

\begin{figure}
\centering
\includegraphics[width = 0.65\linewidth]{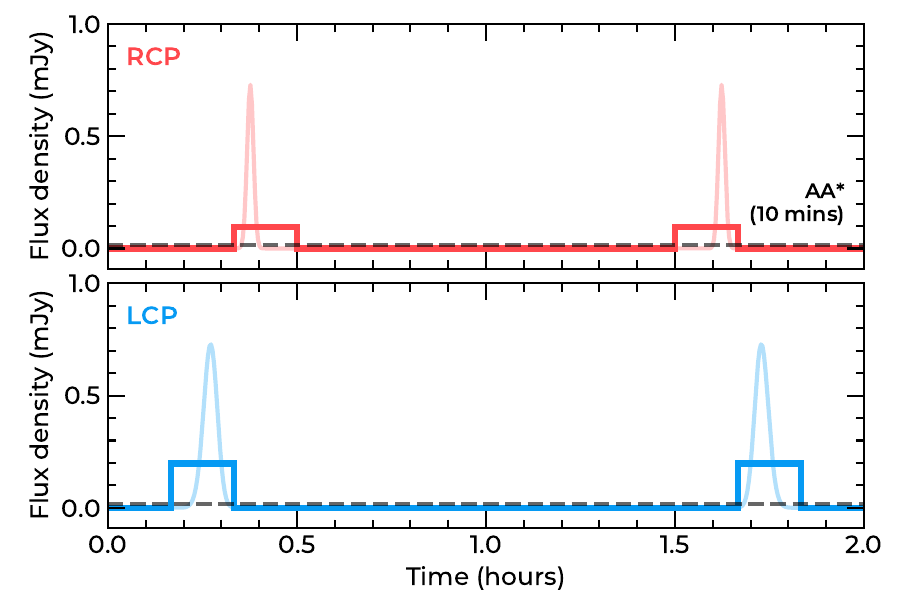}
\caption{Example detection of auroral radio emission at 1~GHz from UCD at 12~pc with AA*. The top and bottom panels show the right and left circularly polarised signal respectively (RCP and LCP). The lighter thinner line in each panel show the auroral signal at a very high time resolution, and the darker stepped line shows the signal binned to a 10 minute resolution. The dashed horizontal line shows the sensitivity of AA* at 1~GHz for a 10 minute exposure computed as described in Section~\ref{sec:EP aurora yields}. At this resolution, the pulses reach SNRs of $\sim6$ and 12 in RCP and LCP respectively. The parameters used are: $i = 90^\circ$, $\beta = 10^\circ$, $B_0 = 1$~kG, $P_\text{rot} = 2$~hours, $\phi_{\text{rot},0} = 0$, $L_\text{B} = 10$~radii, $\phi_\text{B} = 0$, $\alpha = 75^\circ$, $\Delta\alpha = 1^\circ$, $l = 10^{13}$~erg~s$^{-1}$~Hz$^{-1}$.}
\label{fig:UCD aurora example}
\end{figure}

With the required parameters established, we simulate the lightcurves of $10^6$ UCDs for observing times of 1 and 8 hours. We compute each lightcurve at a resolution of 1 second, and then re-bin to a resolution of 10~minutes. An example is shown in Figure~\ref{fig:UCD aurora example}. We then count the number of 10~minute bins that achieve an SNR > 5 as described in Section~\ref{sec:EP aurora yields}. We note that the AFL model produces both right and left circularly polarised signals (RCP and LCP). If the pulses overlap, they could result in a non-detection in Stokes~V. However, for simplicity we assume that a detection is obtained provided either the RCP or LCP signal is bright enough, irrespective of whether they overlap. We also note that this effect can be circumvented if the observing bandwidth is sufficiently wide \citep{lamy08a}.

\subsection{Expected detection yields of UCD aurorae}
\label{sec:UCD aurora detections}

\begin{figure}
\centering
\includegraphics[width = 0.6\linewidth]{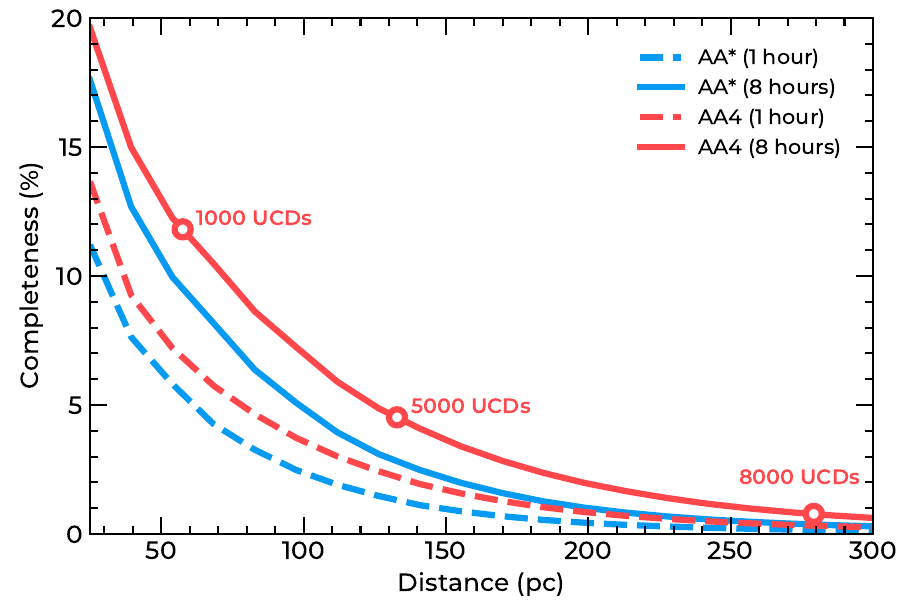}
\caption{The estimated completeness of UCDs as a function of distance with aurorae detectable by SKA-Mid. Results are shown for both the AA* and AA4 configurations, for observing times of 1 and 8 hours. The distances where 1\,000, 5\,000, and 8\,000 UCDs are detected are marked for the 8 hour observations with AA4, assuming the maximum 75\% sky coverage of the SKA is achieved.}
\label{fig:UCD aurora yields}
\end{figure}

In Figure~\ref{fig:UCD aurora yields}, we show the fraction of UCDs with aurorae detectable by SKA-Mid in configurations AA* and AA4 as a function of distance, for observing times of 1 and 8 hours. We estimate that 9.8 to 18.5\% of all UCDs within 25~pc will be detectable with SNRs exceeding 5. The completeness is above 5\% out to 60 to 120~pc, where the known population is very incomplete \citep{best24}, and drops below 1\% at distances of around 150 to 250~pc. It is interesting to note that our estimated completeness within 25~pc is close to the 5-10\% burst rate inferred from volume-limited surveys by \citet{route16b}. This may indicate that most UCDs exhibit radio aurorae, but we only detect a fraction of them due to unfavourable geometry. To further explore this, we re-ran our simulation for $10^5$ UCDs within 25~pc, using a similar observing setup as \citet{route16b} (0.4~mJy noise, 5 GHz observing frequency, SNR detection threshold of 3). From this, we find that $3.3\pm0.5\%$ of the 25~pc sample are detectable. In other words, the occurrence rate of radio bursts on UCDs can be reproduced with our geometric prescription combined with the assumed priors on the magnetic and radio properties of UCDs.

In terms of the actual number of UCDs we expect to be detected with the SKA, the space density of UCDs within 25~pc with spectral types from M7 to M9.5 is estimated to be $7.1\times10^{-3}$~pc$^{-3}$ \citep{bardalezgagliuffi19}, and $8.1\times10^{-3}$~pc$^{-3}$ from L0 to T8 \citep{best24}, giving a total space density of $15.2\times10^{-3}$~pc$^{-3}$. Given our computed completenesses out to 300~pc and the 75\% sky coverage of the SKA, we estimate that 1\,600 to 8\,000 will be detected, which is at least a two order of magnitude increase in the number of UCDs with detected aurorae.

Large radio transient surveys that will be carried out with the SKA will be key to delivering these estimated numbers. Source identification efforts will greatly benefit from the Euclid mission, which is anticipated to detect millions of UCDs within the thin disk of the Milky Way \citep{solano21}. The large statistics the SKA will deliver for radio-emitting UCDs will allow us to quantify the underlying demographics of their magnetic fields and radio emission. This will be particularly fruitful at lower frequencies with SKA-Low, which so far remains largely unexplored for UCDs \citep{zic19, vedantham20, vedantham23, yiu25}. Additionally, our estimated yields here assume a detection is achieved within an 8-hour window. Stacking observations covering multiple rotation phases of UCDs could allow for fainter targets to also be detected.

\subsection{Demographics and detection biases for auroral UCDs}

\begin{figure}
\centering
\includegraphics[width = 0.6\linewidth]{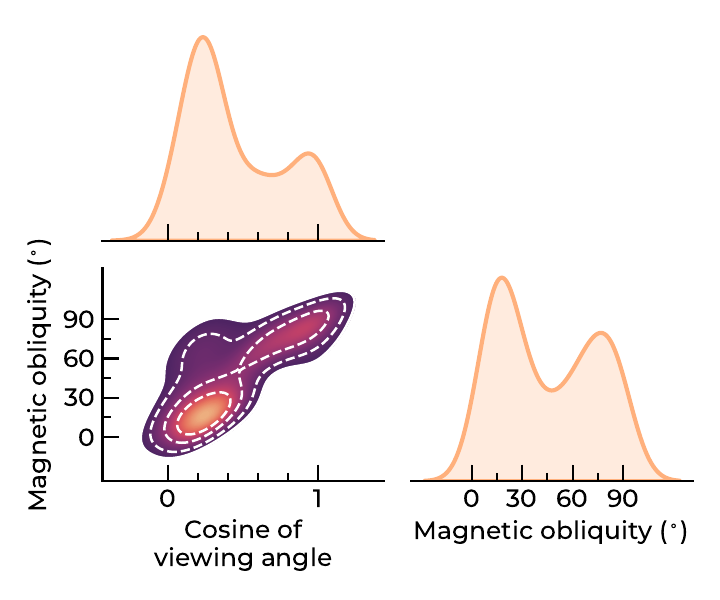}
\caption{The geometric bias we expect for detecting auroral emission on UCDs with the SKA. The filled contour in the bottom left panel shows the bivariate kernal density estimation (KDE) above the 10th percentile for the magnetic obliquity and viewing angle of UCDs with duty cycles exceeding 50\%. The dashed white lines show the 20th, 50th, and 80th percentiles. The two distinct geometric configurations of an edge-on view with a low obliquity and a pole-on view with a high obliquity can be clearly seen. The panels above and to the right show the univariate KDE of each variable. The data used here is that simulated for the 8 hour observations with AA4. Figure made using \texttt{seaborn.PairGrid} \citep{seaborn}.}
\label{fig:UCD aurora geometric bias}
\end{figure}

The large number of UCDs we anticipate the SKA will detect will allow us to explore underlying detection biases. For instance, \citet{kavanagh24} recently suggested that there should be a bias towards detecting radio aurorae on UCDs with certain viewing and magnetic geometries, analogous to the geometric bias expected for detecting magnetic star-planet interactions \citep{kavanagh23}. In Figure~\ref{fig:UCD aurora geometric bias}, we show the viewing angle against the magnetic obliquity of each UCD with a duty cycle exceeding 50\% from the simulated 8 hour observations with AA4. We find a bi-modal distribution akin to that found by \citet{kavanagh23}. The two modes of this distribution describe two distinct geometric configurations for UCDs – an edge-on view with a low obliquity, and a pole-on view with a large obliquity. Both configurations enable the magnetic axis to form the angle $\alpha$ with the line of sight over the UCD's rotation, providing favourable conditions for the emission cone to intersect the line of sight (\citealp[]{kavanagh23}; see also Section 2.3 of \citealp[]{Vedantham01.2026.SKA}).

UCD viewing angles can be inferred from their rotation periods and velocities \citep{vos17}, and their magnetic obliquities can be estimated via \texttt{CHARM} \citet{kavanagh24}. Confirmation of the expected detection bias presented here will improve targeting strategies to boost detection yields. Comparison of the detected distribution of magnetic obliquities will also allow us to estimate the true underlying distribution, which will inform us about magnetic field generation on UCDs, and in turn their interior structures \citep{stanley04}.

In terms of the other parameters, we expect that nearby UCDs with strong magnetic fields, short rotation periods, and bright aurorae are more favourable for detection. Additionally, lower observing frequencies are favoured, as a wider range of magnetic field strengths can be seen. We caution however that the results presented here are dependent on a number of assumptions about the system priors and underlying magnetic geometry. The large statistics of UCDs delivered by the SKA will facilitate extracting the true distributions of radio and magnetic properties for UCDs.

\section{Detecting ultracool dwarf radiation belts with the SKA}

As for their auroral emission, we can use the observed levels of quiescent/radiation belt emission from UCDs to estimate their expected detection yields for the SKA. Additionally, the recent findings from \citet{kao23} and \citet{climent23} have unveiled the prospects for spatially-resolving the radiation belt emission of UCDs using very long baseline interferometry (VLBI). We explore these two scenarios for the SKA in Sections~\ref{sec:UCD belt - unresolved} and \ref{sec:UCD belt - resolved}.


\subsection{Unresolved radiation belts}
\label{sec:UCD belt - unresolved}

The observed unresolved quiescent radio luminosities of individual UCDs range from $\sim4\times10^{10}$ to $4\times10^{14}$~erg~s$^{-1}$~Hz$^{-1}$ (Table~\ref{table:UCD radio detections}), and have been detected from 144~MHz to 100~GHz \citep{williams15c, vedantham20}. While synchrotron emission from a single electron is highly collimated \citep{rybicki86}, radiation belt emission is comprised of synchrotron beams from many individual electrons with different phases, frequencies, and locations within the magnetosphere. As a result, the expected signature should not vary significantly as the UCD rotates, assuming the underlying population of relativistic electrons is stable. This means that detection of UCD radiation belt emission is more suited to single longer exposures, as opposed to the short cadence observations necessary to detect their aurorae.

\begin{figure}
\centering
\includegraphics[width = 0.6\linewidth]{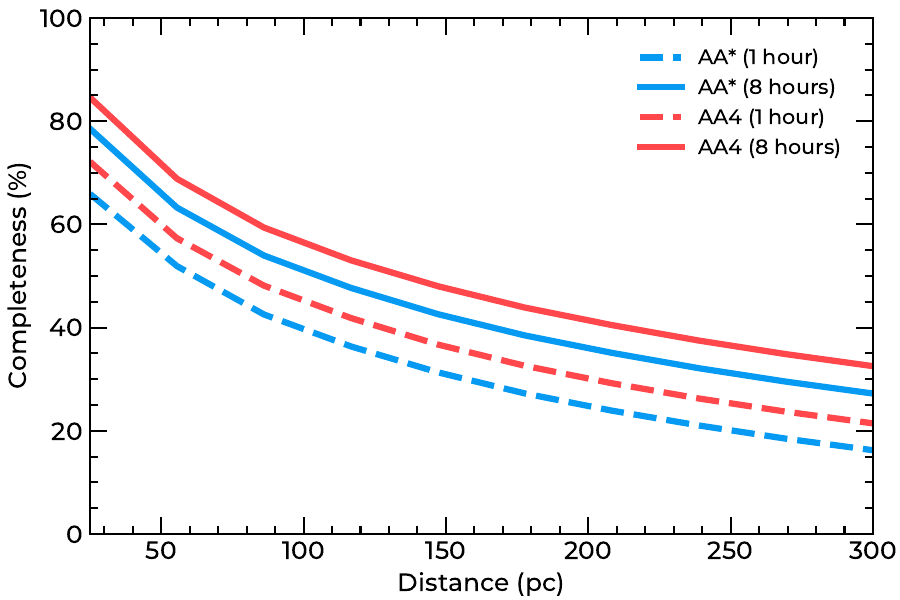}
\caption{Same as Figure~\ref{fig:UCD aurora yields} but for unresolved radiation belt emission from UCDs.}
\label{fig:UCD unresolved belt detections}
\end{figure}

To estimate the number of UCDs with unresolved radiation belts detectable by the SKA, we simulate emission from 1 million UCDs, uniformly sampling the log of the luminosity in the aforementioned range, the distance from 2 to 300~pc as described in Section~\ref{sec:UCD aurora sim}, and the observing frequency from 144~MHz to 1.76~GHz and 4.6 to 15.3~GHz. We then compute the synthetic flux density observed via Equation~\ref{eq:flux luminosity}, and count the number of systems with SNRs of 5 or greater as described in Section~\ref{sec:EP aurora yields}). We show the completeness as a function of distance in Figure~\ref{fig:UCD unresolved belt detections}. We estimate that 66 to 85\% of UCDs within 25~pc are detectable with the SKA. This is significantly larger than the inferred rate of 15 to 20\% for individual UCDs \citep{kao24}. However, we assume that all UCDs possess radiation belt emission, which may not be the case. This contrasts with our findings in Section~\ref{sec:UCD aurora detections}, which suggest that most UCDs possess auroral emission. Additionally, we assume that the lower limit on the luminosity is the lowest observed value, which skews our simulation towards bright UCDs only. Reducing the lower limit of the radio luminosity by three orders of magnitude, we obtain completeness rates of 40 to 50\%. As for their auroral emission, the large number of UCDs with quiescent emission detected with the SKA will provide more robust estimates for these underlying distributions. We also note that \citet{kao25} recently showed that radiation belt emission is enhanced for UCDs in binaries. While the underlying cause of this enhancement remains unclear, this phenomenon can be exploited in the SKA era to achieve higher detection yields by targeting known binaries.

\subsection{Resolvable belts around UCDs}
\label{sec:UCD belt - resolved}

SKA-Mid operating as a phased array within a global VLBI network will deliver sub-milliarcsecond (mas) resolution at GHz frequencies with an unprecedented sensitivity. This will provide the opportunity to spatially-resolve the radiation belt emission of nearby UCDs, a technique recently demonstrated for the UCD LSR~J1835+3259 by \citet{kao23} and \citet{climent23}. These images contain a wealth of information about UCDs, such as the large-scale magnetic field geometry and the energetic electron properties throughout their magnetospheres. Extracting this information will facilitate a better understanding of radiation belt emission detected from more distant UCDs, where spatially resolving the belt is no longer feasible. Temporal tracking of the radiation belt can also provide a direct measurement of the rotation periods of UCDs, which is generally only derivable if the UCD exhibits detectable photometric variability \citep{milespaez23}.

Forming a 10\,000 km baseline with a 100-m dish, we expect that SKA-Mid will achieve a sensitivity at GHz frequencies of 4~µJy per beam for a 1~hour integration \citep{rioja20} and a spatial resolution of 1~mas. The spatial extent of the radiation belt around LSR~J1835+3259 is $\sim35$~Jupiter radii \citep{climent23}, which we expect to be resolvable out to $\sim17$~pc. From our calculations in Section~\ref{sec:UCD belt - unresolved}, we estimate that 78\% of the UCDs within 17~pc have radiation belts with SNRs > 5. This corresponds to around 300 UCDs, assuming they all possess radiation belts.


\section{Detection prospects via astrometric monitoring}

So far, we have focussed on methods for directly detecting and characterising exoplanets and UCDs with the SKA. However, there are also viable pathways to indirectly study these objects. For instance, the SKA is ideally suited to detect magnetic interactions between exoplanets/UCDs with magnetised hosts \citep[see][]{Vedantham01.2026.SKA}. Additionally, exoplanets and UCDs that orbit a radio-emitting star or UCD can induce astrometric motion in the radio centroid of the host which could be resolvable using the SKA in tandem with VLBI. This has already been demonstrated as a viable detection pathway for satellites using current radio telescopes \citep[e.g.][]{forbrich13, curiel20, curiel22}. Given the order of magnitude increase in sensitivity expected for sub-mas imaging with SKA-VLBI, long-term monitoring may unveil a wealth of previously undetected satellites to nearby radio-emitting UCDs and stars.

\begin{figure}
\centering
\includegraphics[width = 0.7\linewidth]{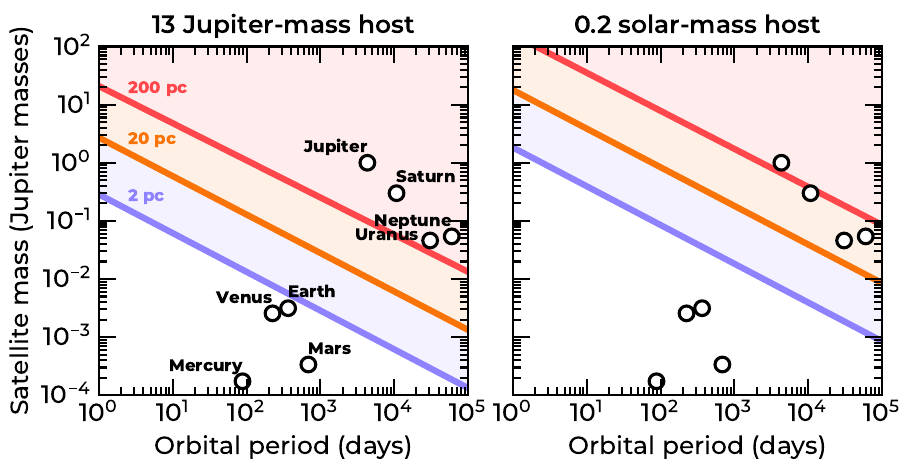}
\caption{The parameter space of satellites around a radio-emitting host UCD (left) and an M~dwarf star (right) that are detectable via astrometric monitoring with SKA-VLBI. The coloured lines show the detection limits at different distances, above which the satellite-induced astrometric motion exceeds 0.5~mas. The Solar System planets are overplotted for reference. Prospects for detecting exoplanets around radio-emitting UCDs are particularly favourable, with planets of a few Earth masses being detectable in orbits less than 1000 days for the closest systems.}
\label{fig:astrometry}
\end{figure}

The angular shift or reflex motion induced by a satellite on a host is \citep{forbrich13}:
\begin{equation}
\theta = \frac{r_\text{orb}}{d} \frac{M_\text{sat}}{M_\text{host}} ,
\end{equation}
where $r_\text{orb}$ is the orbital distance, $d$ is the distance to the system, and $M_\text{sat}$ and $M_\text{host}$ are the masses of the satellite and host. At GHz frequencies, the $5\sigma$ astrometric accuracy of SKA-VLBI is estimated to be $\sim0.5$~mas \citep{rioja20}. In Figure~\ref{fig:astrometry}, we show the parameter space of satellites corresponding to a reflex motion at this level. We show results for both a UCD host of 13 Jupiter masses (the lower limit for UCDs), and an M~dwarf host of 0.2 solar masses, which are the most prevalent radio emitters on the main sequence \citep{driessen24}. In the UCD host case, a 10~Earth-mass planet in a 1000~day orbit is detectable within 22~pc. Prospects for detecting such planets are likely most favourable for UCDs with high duty cycles, such as those with persistent radiation belt emission. From our calculations in Section~\ref{sec:UCD belt - unresolved}, we estimate that $\sim680$ UCDs are within 22~pc with radiation belts at a level detectable with SKA-VLBI (i.e. with fluxes above 40~µJy), again assuming they all possess radiation belts.

These results show the promise of SKA-VLBI for expanding the demographics of satellites around low-mass stars and UCDs. While formation theory predicts that UCDs likely host rocky planets \citep{liu20}, the intrinsic faintness of UCDs hinders their detectability with current technology. Astrometric detection of satellites is also biased towards face-on orbits \citep{smith24, elbadry24}, which cannot be detected via the transit or radial velocity methods. Radio monitoring of such systems may also unveil signatures of magnetic interactions between the host and satellite, which have favourable detection prospects in face-on systems \citep{kavanagh23}.

\section{Looking towards operations}

The uncertainties on the physical and geometric properties of exoplanets and ultracool dwarfs that underpin the brightness of their radio emission in frequency and time mean that their detection prospects for the SKA are most favourable for wide-field surveys with high time cadences. One approach could be to monitor specific regions of the sky via regular SKA Cycle proposals. Alternatively, wide field transient surveys akin to those being carried out with LOFAR will likely yield a large number of detections \citep{shimwell22, callingham23}. This however will be a significant undertaking, and is likely at a level appropriate for a Key Science Project. Collaboration and efficient communication between different working groups will be key to maximising the science output for all in this case.

Targeted observations will also be a viable detection strategy in specific cases, such as for monitoring the most promising exoplanetary systems \citep[e.g.][]{turner21, zhang25} and the known population of radio-emitting UCDs. New candidate systems detected in surveys with the SKA and other operating radio telescopes may also be suited to targeted follow-up. Detecting satellites around nearby radio-emitting stars and UCDs through astrometry will also demand targeted monitoring over multiple years. Any one of these approaches will likely deliver unprecedented insights into the formation and evolution of extrasolar worlds.


\section*{Acknowledgements}

We thank the anonymous reviewer for their time and comments, which strengthened the cohesion of the chapter. RDK and JRC acknowledge funding from the European Union via the European Research Council (ERC) grant Epaphus (project number: 101166008). PZ and CKL acknowledge funding from the ERC under the European Union Horizon 2020 research and innovation programme (grant agreement no. 101020459 - Exoradio). RDK and HKV acknowledge funding from the Dutch Research Council (NWO) for the e-MAPS (Exploring Magnetism on the Planetary Scale) project (project number VI.Vidi.203.093) under the NWO talent scheme Vidi. HKV also acknowledges funding from the European Research Council under the European Union’s Horizon Europe programme (grant number 101042416 STORMCHASER). AZ acknowledges support from ANID -- Millennium Science Initiative Program -- Center Code NCN2024\_001 and Fondecyt Regular grant number 1250249.


\newcommand{\actaa}{Acta Astron.} 
\newcommand{\araa}{ARA\&A} 
\newcommand{\aar}{A\&ARv} 
\newcommand{\aapr}{A\&ARv} 
\newcommand{\ab}{Astrobiol.} 
\newcommand{\aj}{AJ} 
\newcommand{\apj}{ApJ} 
\newcommand{\apjl}{ApJL} 
\newcommand{\apjs}{ApJSS} 
\newcommand{\ao}{Appl. Opt.} 
\newcommand{\apss}{Astro. \& Space Sci.} 
\newcommand{\aap}{A\&A} 
\newcommand{\aaps}{A\&AS.} 
\newcommand{\baas}{Bull. Am. Astron. Soc.} 
\newcommand{\caa}{Chinese A\&A} 
\newcommand{\cjaa}{Chinese J. A\&A} 
\newcommand{\cqg}{Class. Quantum Gravity} 
\newcommand{\gal}{Galaxies} 
\newcommand{\gca}{Geo. Cosmo. Acta} 
\newcommand{\grl}{Geophys. Res. Let.}
\newcommand{\icarus}{Icarus} 
\newcommand{\jcap}{JCAP} 
\newcommand{\jgr}{J. Geophys. Res.} 
\newcommand{\jgrp}{J. Geophys. Res. Planets} 
\newcommand{\jqsrt}{J. Quant. Spectrosc. Radiat. Transf.} 
\newcommand{\memsai}{Mem. SAIt} 
\newcommand{\mnras}{MNRAS} 
\newcommand{\nat}{Nature} 
\newcommand{\nastro}{Nat. Astron.} 
\newcommand{\ncomms}{Nat. Commun.} 
\newcommand{\nphys}{Nat. Phys.} 
\newcommand{\na}{New Astron.} 
\newcommand{\nar}{New Astron. Rev.} 
\newcommand{\physrep}{Phys. Rep.} 
\newcommand{\pra}{Phys. Rev. A} 
\newcommand{\prb}{Phys. Rev. B} 
\newcommand{\prc}{Phys. Rev. C} 
\newcommand{\prd}{Phys. Rev. D} 
\newcommand{\pre}{Phys. Rev. E} 
\newcommand{\prx}{Phys. Rev. X} 
\newcommand{\prl}{Phys. Rev. Let.} 
\newcommand{\psj}{Planet. Sci. J.} 
\newcommand{\planss}{Planet. Space Sci.} 
\newcommand{\pnas}{Proc. Natl Acad. Sci. USA} 
\newcommand{\procspie}{Proc. SPIE} 
\newcommand{\pasa}{PASA} 
\newcommand{\pasj}{PASJ} 
\newcommand{\pasp}{PASP} 
\newcommand{\rmxaa}{RMXAA} 
\newcommand{\sci}{Science} 
\newcommand{\sciadv}{Sci. Adv.} 
\newcommand{\solphys}{Sol. Phys.} 
\newcommand{\sovast}{Soviet Ast.} 
\newcommand{\ssr}{Space Sci. Rev.} 
\newcommand{\uni}{Universe} 

\bibliographystyle{abbrvnat-maxbibnames4}
\bibliography{bibliography} 

\end{document}